# Semi-quantum private comparison protocol under an almost-dishonest third party

**Wen-Han Chou[1], Tzonelih Hwang[*] and Jun Gu[2]**

*Department of Computer Science and Information Engineering, National Cheng Kung University, No. 1, University Rd., Tainan City, 70101, Taiwan, R.O.C.*

[1] a29554483@gmail.com

[*] hwangtl@ismail.csie.ncku.edu.tw

[2] isgujun@gmail.com

[*]**Responsible for correspondence:**

Tzonelih Hwang

Distinguished Professor

Department of Computer Science and Information Engineering,

National Cheng Kung University,

No. 1, University Rd.,

Tainan City, 70101, Taiwan, R.O.C.

Email: hwangtl@ismail.csie.ncku.edu.tw

TEL: +886-6-2757575 ext. 62524

**Abstract**

This study presents the first semi-quantum private comparison (SQPC) protocol under an almost-dishonest third party. The proposed protocol allows two classical participants to compare their secret information without compromising it's privacy. The security analyses indicate that the protocol is free from several well-known attacks.

**1 Introduction**

Since the first quantum key distribution (QKD) protocol presented by Bennett and Brassard [1], many quantum cryptographic protocols, such as teleportation [2-7], quantum secret sharing (QSS) [8-16], quantum secure direct communication (QSDC) [17-21], and so on, have been proposed to solve various security problems. Recently, quantum private comparison (QPC) has become another popular topic for research. The goal of a QPC protocol is to privately compare two participants' undisclosed information for equality. In previous QPC protocols, the participants need to be equipped with advanced quantum devices such as quantum memory, quantum generator, or quantum unitary operations for comparison.

However, this paper proposes the first semi-quantum private comparison (SQPC) protocol for the participants who do not have any quantum capability to compare their secrets under an **almost-dishonest third-party** [22, 23]. An almost-dishonest TP may try to perform various possible attacks for disclosing the secret information of the participants, but he/she cannot collude with the participants.

In previous, Boyer et al. [24, 25] proposed two novel semi-quantum key distribution (SQKD) protocols using single photons. According to their definition, the term "semi-quantum" implies that the sender, Alice, is a powerful quantum communicant, whereas the receiver, Bob, has only classical capabilities. More precisely, the sender (Alice) has the ability to perform the following operations: (1) prepare any quantum state, such as single photons or Bell states, (2) perform any quantum measurement, such as Bell measurement or multi-qubit joint measurement, and (3) store qubits in a quantum memory. Conversely, the classical Bob is restricted to perform the following operations over the quantum channel: (1) prepare new qubits in the classical basis $\{|0\rangle, |1\rangle\}$ (i.e., Z basis), (2) measure qubits in the classical basis, (3) reorder the qubits via different delay lines, and (4) send or reflect the qubits without disturbance.

As the classical basis only considers the qubits $|0\rangle$ and $|1\rangle$, the other quantum superpositions of single photons are not assumed for the classical Bob. The classical Bob's operations described above are equivalent to the traditional $\{0,1\}$ computation. In this paper, we propose an SQPC protocol, where the classical participants are limited to perform the operations (1), (2), and (4). That is, the participants do not have any quantum capabilities, whereas the TP is the only one who has the quantum capability. A semi-quantum private comparison protocol can reduce not only the computational burden of the communicants but also the cost of the quantum hardware devices in practical implementations. Furthermore, the proposed protocols are free from various well-known attacks with the assumption of only an almost dishonest third party.

The rest of this paper is organized as follows: Section 2 proposes the semi-quantum private comparison (SQPC) protocol. Section 3 presents security analyse of the proposed SQPC protocol. Section 4 summarizes our results.

## 2 Proposed SQPC protocol

This section proposes a new semi-quantum private comparison protocol under an almost-dishonest third party. Alice and Bob are two participants who want to compare the equality of their secret information in the SQPC protocol. TP is assumed to be an almost-dishonest third party who may try to derive information during executing of the protocol. TP may modify the procedure of the protocol to derive information, but he cannot publish a fake comparison result or collaborate with any client. The quantum channels are assumed to be ideal (i.e., non-lossy and noiseless), and the public channels between TP and Alice, TP and Bob, and Alice and Bob are also assumed to be authenticated. Based on the three-party scenario described above, the process of the proposed protocol will be described in steps as follows.

### 2.1 Description of the proposed SQPC protocol

Before presenting the steps of the proposed protocol, we briefly describe some rules for malicious behavior detection, some symbols, and entanglement swapping used in our protocol. In particular, for each received qubit, the participant has to set the values of $l_1$ and $l_2$ for that qubit, respectively. $l_1=0/1$ denotes that the participant selects the qubit for malicious behavior detection : $l_1=0$ indicates that the qubit is used to detect attacks from any malicious users, whereas $l_1=1$ is used to detect attacks from TP. $I_2$ indicates the type of operations performed on that qubit by the client. $I_2=0$ represents that the participant performs the Z-basis measurement and $I_2=1$ represents a direct reflection of the bit to TP by the client.

The entanglement swapping allows one to measure on any two independent entangled photons, the rest photons will be entangled regardless of the distance between them. For example, assume that there are two independent EPR $|\phi^+\rangle$ pairs, namely the pair 1-2 and the pair 3-4. Then if we perform Bell measurement on the qubit 1 and the qubit 3, then the qubit 2 and the qubit 4 which are originally independent will be entangled immediately. If the measurement result of the qubit 1 and the qubit 3 is $|\phi^-\rangle$, then the state of the qubit 2 and the qubit 4 must be $|\phi^-\rangle$ as well. We may encode the Bell measurement result in the classical bits as follows: $|\phi^+\rangle = 00$, $|\phi^-\rangle = 01$, $|\varphi^+\rangle = 10$, and $|\varphi^-\rangle = 11$. The entanglement swapping follows the math formula $M_{Bell}q_1q_2 \oplus M_{Bell}q_3q_4 = M_{Bell}q_1q_3 \oplus M_{Bell}q_2q_4$ [26, 27], where $\oplus$ is the exclusive-OR operation. In our scheme, TP will set a value of M according to this math formula. M=0 represents the measurement results and the math formula match, whereas M=1 represents the measurement results and the math formula do not match.

Based on the coordination of the values of $(I_1, I_2)$ and M, the proposed SQPC allows two mutually untrusted participants to compare the equality of their secrets under the almost dishonest TP. The detail of the protocol is described in steps as follows:

**Step1.** TP randomly generates two sequences of Bell states $S_A = \{S_{A_1}, S_{A_2}, S_{A_3}, ..., S_{A_n}\}$, $S_B = \{S_{B_1}, S_{B_2}, S_{B_3}, ..., S_{B_n}\}$ with the length of n, where $S_{A_i} = \{a_1^i, a_2^i\}$, $S_{B_i} = \{b_1^i, b_2^i\}$ for i =1,2,…,n. The initial state of each pair in $S_{A_i}$ ($S_{B_i}$) is denoted as $IS_{A_i}$ ($IS_{B_i}$) ($IS_{A_i}$ and $IS_{B_i} \in \{|\phi^+\rangle, |\phi^-\rangle, |\varphi^+\rangle, |\varphi^-\rangle\}$). Then, TP divides these two sequences $S_A$ and $S_B$ into four ordered sequences, $A_1 = \{a_1^i\}$, $A_2 = \{a_2^i\}$, $B_1 = \{b_1^i\}$ and $B_2 = \{b_2^i\}$ for i =1,2,…,n, which denote all the first and the second particles of Bell states in $S_A$ and $S_B$, respectively.

**Step2.** After the **Step1's** preparation, TP retains the sequences $A_2$ and $B_2$, and sends a particle in the sequences $A_1$ to Alice and a particle in $B_1$ to Bob respectively.

**Step3.** When Alice (Bob) receives a qubit $a_1^i$ ($b_1^i$) in $A_1$ ($B_1$), she (he) has to set the values of $I_1^i$ and $I_2^i$ for that qubit. If $I_1^i=0$, Alice (Bob) will send $I_1^i$ and $I_2^i$ back to TP, whereas when $I_1^i=1$, Alice (Bob) will only send $I_1^i$ back to TP and keep $I_2^i$ in her (his) hand. TP will perform three different operations according to $I_1^i$ and $I_2^i$ as shown in **Table1**. In our scheme, if a participant wants to check Eve's attack, he/she will set $l_1=0$. If a participant wants to check the honesty of TP, then he/she sets $l_1=1$. TP can only be checked when both participants want to do so, i.e., $l_1=1$ should be set by both clients for that qubit. If only one participant wants to check TP, then he/she will be ignored. The protocol can be described in the following three cases.

**Step4**.

**Case1.** If Alice (Bob) sets $I_1^i = 0$ and $I_2^i=0$, then she (he) first sends the value of $I_1^i$ and the measured qubit to TP. Because Alice (Bob) might change the state of the photon which comes from TP, we use $a_1^{*i}$ ($b_1^{*i}$) to denote the photon which Alice (Bob) sends back to TP. After TP receives $I_1^i=0$ and $a_1^{*i}$, he publicly announces an acknowledgment. When Alice (Bob) receives the acknowledgment, Alice (Bob) sends $I_2^i=0$ and the measurement result $MR_{a_1^i}$ ($MR_{b_1^i}$) to TP. When TP receives the measurement result $MR_{a_1^i}$ ($MR_{b_1^i}$) and $I_2^i=0$, TP performs a Z-basis measurement on $a_1^{*i}$ ($b_1^{*i}$) and obtains the measurement result $MR_{a_1^{*i}}$ ($MR_{b_1^{*i}}$). Then TP compares the measurement result $MR_{a_1^i}$ ($MR_{b_1^i}$) with the measurement result $MR_{a_1^{*i}}$ ($MR_{b_1^{*i}}$) to check the existence of an eavesdropper. If the measurement results are not equal, then TP will terminate the protocol and start from the beginning. If the measurement results are equal, then TP will follow the **Step2** to send out the next qubits to the clients.

**Case2.** If Alice (Bob) sets $I_1^i=0$ and $I_2^i=1$, then Alice (Bob) first sends the original qubit and $I_1^i=0$ to TP. After TP receives the qubit and $I_1^i=0$, he publicly announces an acknowledgment. After Alice (Bob) receives the acknowledgment, she (he) sends $I_2^i=1$ to TP. TP then performs a Bell measurement on the qubit $a_1^{*i} a_2^i$ ($b_1^{*i} b_2^i$) and compares the Bell measurement result with the initial state of $S_{A_i}$ ($S_{B_i}$) to see whether they are equal or not. If the measurement results are not equal, then TP will terminate the protocol and start from the beginning. If the measurement results are equal, then TP will follow the **Step2** to send out the next qubits to the clients.

**Case3.** If Alice and Bob both set $I_1^i=1$, then they send the qubits and $I_1^i=1$ to TP. When TP receives both $I_1^i=1$ from Alice and Bob, TP performs two Bell measurements on the qubits $a_1^{*i} b_1^{*i}$ and $a_2^i b_2^i$, which are denoted as $M_{Bell} a_1^{*i} b_1^{*i}$ and $M_{Bell} a_2^i b_2^i$ respectively. Then TP sets a value for the variable M according to the math formula mentioned earlier (i.e., $IS_{A_i} \oplus IS_{B_i} = M_{Bell} a_1^{*i} b_1^{*i} \oplus M_{Bell} a_2^i b_2^i$).

M=1 if $IS_{A_i} \oplus IS_{B_i}$ is not equal to $M_{Bell} a_1^{*i} b_1^{*i} \oplus M_{Bell} a_2^i b_2^i$, whereas M=0 if

$IS_{A_i} \oplus IS_{B_i}$ is equal to $M_{Bell}a_1^{*i}b_1^{*i} \oplus M_{Bell}a_2^i b_2^i$. Then the protocol continues to the next step.

**Table 1.**

| Alice | Bob | TP |
|---|---|---|
| $I_1=0,\ I_2=0$ | $I_1=0,\ I_2=0$ | Z-basis measurement $a^*_1, b^*_1$ |
| $I_1=0,\ I_2=0$ | $I_1=0,\ I_2=1$ | Z-basis measurement $a^*_1$ <br> Bell measurement $(b^*_1, b_2)$ |
| $I_1=0,\ I_2=0$ | $I_1=1,\ I_2=0$ | Z-basis measurement $a^*_1$ <br> Ignore Bob |
| $I_1=0,\ I_2=0$ | $I_1=1,\ I_2=1$ | Z-basis measurement $a^*_1$ <br> Ignore Bob |
| $I_1=0,\ I_2=1$ | $I_1=0,\ I_2=0$ | Bell measurement $(a^*_1, a_2)$ <br> Z-basis measurement $b^*_1$ |
| $I_1=0,\ I_2=1$ | $I_1=0,\ I_2=1$ | Bell measurement $(a^*_1, a_2)$ <br> Bell measurement $(b^*_1, b_2)$ |
| $I_1=0,\ I_2=1$ | $I_1=1,\ I_2=0$ | Bell measurement $(a^*_1, a_2)$ <br> Ignore Bob |
| $I_1=0,\ I_2=1$ | $I_1=1,\ I_2=1$ | Bell measurement $(a^*_1, a_2)$ <br> Ignore Bob |
| $I_1=1,\ I_2=0$ | $I_1=0,\ I_2=0$ | Ignore Alice <br> Z-basis measurement $b^*_1$ |
| $I_1=1$, $I_2=0$ | $I_1=0$, $I_2=1$ | Ignore Alice |
| | | Bell measurement ($b^*_1, b_2$) |
| $I_1=1$, $I_2=0$ | $I_1=1$, $I_2=0$ | Bell measurement ($a^*_1, b^*_1$) |
| | | Bell measurement ($a_2, b_2$) |
| $I_1=1$, $I_2=0$ | $I_1=1$, $I_2=1$ | Bell measurement ($a^*_1, b^*_1$) |
| | | Bell measurement ($a_2, b_2$) |
| $I_1=1$, $I_2=1$ | $I_1=0$, $I_2=0$ | Ignore Alice |
| | | Z-basis measurement $b^*_1$ |
| $I_1=1$, $I_2=1$ | $I_1=0$, $I_2=1$ | Ignore Alice |
| | | Bell measurement ($b^*_1, b_2$) |
| $I_1=1$, $I_2=1$ | $I_1=1$, $I_2=0$ | Bell measurement ($a^*_1, b^*_1$) |
| | | Bell measurement ($a_2, b_2$) |
| $I_1=1$, $I_2=1$ | $I_1=1$, $I_2=1$ | Bell measurement ($a^*_1, b^*_1$) |
| | | Bell measurement ($a_2, b_2$) |

**Step5.** If M = 0, TP has to send M and the Bell measurement result on the qubits $a^{*i}_1 b^{*i}_1$, i.e., $|\phi^+\rangle = 00$, $|\phi^-\rangle = 01$, $|\varphi^+\rangle = 10$, or $|\varphi^-\rangle = 11$ to Alice and Bob. After Alice and Bob receive these M and the Bell measurement result, they will perform public discussion based on the values of $I^i_2$. If Alice and Bob both set $I^i_2 = 0$ in **Step3**, then they have to send the Z-basis measurement result on that qubit (i.e., $MR_{a^i_1}$ and $MR_{b^i_1}$) to each other via an authenticated channel. Then Alice and Bob

both compare $MR_{a_1^i}$ and $MR_{b_1^i}$ based on $M_{Bell}a_1^{*i}b_1^{*i}$. That is: if $M_{Bell}a_1^{*i}b_1^{*i}=\left|\phi^{\pm}\right\rangle$, then $MR_{a_1^i}$ and $MR_{b_1^i}$ must be either $|0\rangle,|0\rangle$ or $|1\rangle,|1\rangle$. If $M_{Bell}a_1^{*i}b_1^{*i}=\left|\varphi^{\pm}\right\rangle$, then $MR_{a_1^i}$ and $MR_{b_1^i}$ must be either $|0\rangle,|1\rangle$ or $|1\rangle,|0\rangle$. If the measurement results do not match then it implies TP does not honestly follow the procedure of the protocol. Hence, Alice and Bob will terminate the protocol and start from the beginning. If the measurement results match, the protocol will go back to the **Step2** for the next qubit. For the other cases, i.e., Alice and Bob both set $I_2^i = 1$ or one sets $I_2^i = 1$ and the other sets $I_2^i = 0$, then it implies either TP honestly follow the procedure of the protocol or the clients do not have enough information to check the honesty of TP. Hence, the protocol will go back to the **Step2** for the next qubit.

On the other hand, if M = 1, then TP has to send M to Alice and Bob. Hereafter, Alice and Bob will perform public discussion based on the value of $I_2^i$. If Alice and Bob both set $I_2^i=1$ in **Step3**, then it implies TP does not honestly follow the procedure of the protocol. Hence, they will terminate the protocol and start from the beginning. For the other cases, i.e., one client sets $I_2^i = 1$ and the other client sets $I_2^i = 0$, then it implies that the clients do not have enough information to check the honesty of TP. The protocol will go back to the **Step2** for the next qubit. Only when Alice and Bob both set $I_2^i=0$, the protocol will continue to the next step.

**Step6.** Up to this step, we know that both Alice and Bob have set $I_1^i = 1$ and $I_2^i = 0$ and have received M=1 from TP. In this step, Alice and Bob will use their measurement results to compare their secret messages. Let, $M_A^i$ and $M_B^i$ be the ith bit of secret messages of Alice and Bob respectively. Alice and Bob individually compute $R_A^i = MR_{a_1^i} \oplus M_A^i$ and $R_B^i = MR_{b_1^i} \oplus M_B^i$ and then send $R_A^i$ and $R_B^i$ to TP.

Upon receiving $R_A^i$ and $R_B^i$, TP transforms the Bell measurement result $a_1^{*i} b_1^{*i}$ into a classical bit string $C_T^i$ ($|\phi^+\rangle = 0$, $|\phi^-\rangle = 0$, $|\varphi^+\rangle = 1$, $|\varphi^-\rangle = 1$) and calculates the comparison results $R^i = R_A^i \oplus R_B^i \oplus C_T^i$. If there is a bit '1' in R, then TP terminates the protocol, and publishes '1' indicating that Alice's and Bob's information is different. Otherwise, TP repeats the protocol from **Step 2** to **Step 6** until all the secret messages are compared. If there are all zeros in R, then TP announces that the two participants' information are identical.

## 3 Security analyses

To analyze the security of the proposed protocol, this section is divided into two parts to focus on two different attacks (the outsider attack and the insider attack). The proposed protocol contains three insiders (i.e., an almost-dishonest third party, TP, and two participants: Alice and Bob). While Section 3.1 investigates the outside eavesdropper's attack, Section 3.2 analyzes the possibility for an insider (i.e., TP or a participant) to obtain the other participant's secret information.

### 3.1 Outsider attack

A malicious outsider, Eve, may try to perform some well-known attacks, such as the intercept-and resend attack [28] by intercepting the qubit sequences $A_1$ and $B_1$ transmitted from TP to Alice and Bob respectively in **Step2** and then measuring the qubits in the sequences. However, since Eve does not know the value of $I_2^i = 0/1$, if Eve measures on the transmitted sequences, she will be detected with a probability of $1 - (\frac{7}{8})^n$ in **Step4,** where n denotes the number of qubits in the sequences. If n is large enough, the probability will be close to one. This is because: Alice (Bob) has $\frac{1}{2}$ probability to set $I_1^i=0$ and $\frac{1}{2}$ probability to set $I_2^i=1$ and TP has $\frac{1}{2}$ probability to get the different result from $IS_{A_i}$ ($IS_{B_i}$). Hence, Eve has $\frac{7}{8}$ probability to pass the detection for each qubit. Hence, with n qubits, Eve will be detected with a probability of $1-(\frac{7}{8})^n$ in **Step4.** For example: assume the initial state $S_{A_1^i}=|\phi^+\rangle$ ($S_{B_1^i}=|\phi^+\rangle$), Eve intercepts $a_1^i$ ($b_1^i$) and measures $a_1^i$ ($b_1^i$) with Z basis to get $|0\rangle$, and sends $|0\rangle$ to Alice (Bob). Alice (Bob) sets $I_1^i=0$ and $I_2^i=1$ and sends the qubit $a_1^{*i}$ ($b_1^{*i}$) to TP. After TP receives $a_1^{*i}$ ($b_1^{*i}$) from Alice (Bob), Alice (Bob) sends $I_2^i$ to TP. Because $I_1^i=0$ and $I_2^i=1$, TP performs Bell measurement on $a_1^{*i}a_2^i$ ($b_1^{*i}b_2^i$). TP has a probability of $\frac{1}{2}$ to get the measurement result $|\phi^-\rangle$. If TP gets the different result from $IS_{A_i}$ ($IS_{B_i}$), he/she knows that there is an eavesdropper.

In some cases, Eve might intercept the qubit sequences $A_1$ and $B_1$ transmitted from TP to Alice and Bob respectively in **Step2** and send the fake qubits to Alice and Bob. After Alice and Bob receive the fake qubits and set the values of $I_1^i$ and $I_2^i$, they send the fake qubits to TP. Then Eve intercepts the fake qubits and resends the previous qubit sequences $A_1$ and $B_1$ to TP. However, since Eve does not know the value of $I_2^i$ (=0/1), if Eve sends the fake qubits to Alice and Bob, she will be detected with a probability of $1-(\frac{7}{8})^n$ in **Step4,** where n denotes the number of qubits in the sequences. Since Alice (Bob) has $\frac{1}{2}$ probability to set $I_1^i=0$ and $\frac{1}{2}$ probability to set $I_2^i=0$ and TP has $\frac{1}{2}$ probability to get the different result from $M_za_1^{*i}$ ($M_zb_1^{*i}$) and the measurement result which Alice (Bob) sends, Eve has $\frac{7}{8}$ probability to pass the detection for each qubit. Hence, with n qubits, Eve will be detected with a probability of $1-(\frac{7}{8})^n$ in **Step4.** For example: assume that the initial state $S_{A_1^i} = |\phi^+\rangle$ ($S_{B_1^i} = |\phi^+\rangle$). Eve intercepts $a_1^i$ ($b_1^i$) and sends a fake qubit $|0\rangle$ to Alice (Bob). If Alice (Bob) sets $I_1^i = 0$ and $I_2^i = 0$, then Alice (Bob) performs Z-basis measurement on Eve's qubit and sends $a_1^{*i}$ ($b_1^{*i}$) to TP. Eve intercepts $a_1^{*i}$ ($b_1^{*i}$) and resends $a_1^i$ ($b_1^i$) to TP. After TP receives $a_1^i$ ($b_1^i$) from Eve, Alice (Bob) sends $I_2^i$ and the measurement result of Eve's qubit to TP. Because $I_1^i = 0$ and $I_2^i = 0$, TP performs Z-basis measurement on $a_1^{*i}$ ($b_1^{*i}$). Then TP has a probability of $\frac{1}{2}$ to get the measurement result $|1\rangle$. If $M_z a_1^{*i}$ ($M_z b_1^{*i}$) is different from the measurement result which Alice (Bob) sends, then TP will detect the existence of an eavesdropper. Consequently, the outside attacker cannot get any useful information from the participants without being detected.

**3.2 Insider attack**

In this sub-section, two cases of insider attacks are considered. The first case discusses the possibility for a participant to obtain the other participant's secret information. The second case discusses the possibility for TP to retrieve the two participants' secret information.

**Case 1** Participant attack

Suppose Alice is a malicious participant who tries to reveal the other participant's (Bob) secret information. If Alice tries to intercept the transmitted photons $b_1^i$ from TP to Bob, she will be caught as an outside attacker as described in Sect. 3.1. Because when Bob sets $I_1^i=0$, Alice cannot involve in the discussion of TP and Bob. So if Alice wants to reveal Bob's information $M_B^i$, she will be detected by TP like an Eve and the detection rate is the same as described in Sect. 3.1. Similarly, if Bob tries to perform an attack on the protocol, he will be detected too.

**Case 2** TP attack

Because TP is an almost-dishonest third party, he might not faithfully execute the procedure of the protocol, and try to reveal participant's information. When $I_1^i=0$, we detect either Eve's attack or a participant's attack with TP's help. On the contrary we detect TP's attack when Alice and Bob both set $I_1^i=1$. If TP wants to reveal a participant's secret information, he/she can generate fake initial states like Z-basis single photons instead of Bell state. When Alice and Bob measure the qubit with Z-basis and publish $R_A^i$ and $R_B^i$, TP can directly learn Alice's and Bob's secret information. However, in our proposed protocol, Alice and Bob have to discuss the entanglement of $M_{Bell}a_1^{*i}b_1^{*i}$ and $M_{Bell}a_2^i b_2^i$ in **Step5**. If TP generates Z-basis single photons, he does not know the entanglement of $M_{Bell}a_1^{*i}b_1^{*i}$ and $M_{Bell}a_2^i b_2^i$. Hence, he does not know how to set the value of M. TP will be detected in **Step5** when participants both set $I_1^i=1$ and $I_2^i=1$ and the detection rate is about $1-(\frac{15}{16})^n$. If n is large enough, the probability will be close to one. This is because: Alice and Bob both have $\frac{1}{2}$ probability to set $I_1^i=1$ and $\frac{1}{2}$ probability to set $I_2^i=1$. Hence, TP has a probability of $\frac{15}{16}$ to pass the detection for each qubit. Hence, with n qubits, Eve will be detected with a probability of $1-(\frac{15}{16})^n$ in **Step5.**

TP could generate two Bell states and performs Bell measurement on $b_1^{*i} b_2^i$

instead on $a_1^{*i} b_1^i$. Then TP compares the Bell measurement result with $IS_{B_i}$. If the two results are different then it implies Bob sets $I_1^i=1$ and $I_2^i=0$, TP will perform Z-basis measurement on $a_1^{*i}$ to reveal Alice's secret information and set M=1. If the two results are the same, TP will set M=0. In this attack, however, if TP sets M=0, then TP does not know the Bell measurement result of $a_1^{*i} b_1^{*i}$. TP will be detected in Step5 when participants both set $I_1^i=1$ and $I_2^i=0$ and the detection rate is about $1-(\frac{31}{32})^n$. If n is large enough, the probability will be close to one. This is because: Alice and Bob both have $\frac{1}{2}$ probability to set $I_1^i=1$ and $\frac{1}{2}$ probability to set $I_2^i=0$ and TP has $\frac{1}{2}$ probability to send wrong Bell measurement result of $a_1^{*i} b_1^{*i}$ to Alice and Bob. Hence, TP has a probability of $\frac{31}{32}$ to pass the detection for each qubit. Hence, with n qubits, Eve will be detected with a probability of $1-(\frac{31}{32})^n$ in **Step5.**

In our protocol, only TP has the information about the initial state. Therefore he can reveal the comparison result of both participants, but TP cannot know the participants' secret information.

# 4 Conclusions

This paper presents the first semi-quantum private comparison (SQPC) protocol under an almost-dishonest third-party. A semi-quantum private comparison protocol can reduce not only the computational burden of the communicants but also the cost of the quantum hardware devices in practical implementations. The security analysis shows that the proposed protocols are free from the outsider attack and the insider attack. It should also be noted that, like the other semi-quantum schemes, the proposed protocol also suffers from the Trojan-horse attacks [29-31]. To prevent this kind of attacks, the photon number splitter device and wavelength filter device could be adopted [32, 33]. Throughout this paper, we have assumed an almost–dishonest TP. In the future, it could be very challenging to further reduce the trustworthy news of the TP to an **individually dishonest** as defined in [34], where TP cannot even publish a fake comparison result.

# Acknowledgment

We would like to thank the Ministry of Science and Technology of Republic of China for financial support of this research under Contract No. MOST 104-2221-E-006-102-.

communication based on the secret transmitting order of particles," *Physical Review A,* vol. 74, p. 054302, 2006.

[34] Shih-Min Hung and Tzonelih Hwang, "Multiparty Quantum Private Comparsion with individually dishonest Third Parties for strangers,"*arXiv:1607.07119,* 2016.
19